\documentclass[pdflatex, sn-mathphys-num]{sn-jnl}

\usepackage{graphicx}
\usepackage{multirow}
\usepackage{amsmath,amssymb,amsfonts}
\usepackage{amsthm}
\usepackage{mathrsfs}
\usepackage[title]{appendix}
\usepackage{xcolor}
\usepackage{textcomp}
\usepackage{manyfoot}
\usepackage{listings}
\usepackage{physics}
\usepackage{subcaption}
\usepackage{float}

\title{Relativistic Signatures of Dark Matter Equations of State in Static Spherically Symmetric Spacetimes}
\author[1]{\fnm{Akash} \sur{Yadav}} \email{2024rpy9120@mnit.ac.in}
\author[1]{\fnm{Sudhava} \sur{Yadav}} \email{2021rpy9099@mnit.ac.in}
\author*[1]{\fnm{K.K.} \sur{Venkataratnam}} \email{kvkamma.phy@mnit.ac.in}
\affil[1]{
    \orgdiv{Department of Physics},
    \orgname{Malaviya National Institute of Technology Jaipur} ,
    \orgaddress{J.L.N. Marg},
    \city{Jaipur, Rajasthan}, \postcode{302017}, \country{India}
}

\abstract{
We study how three physically motivated dark matter equations of state alter the spacetime geometry of static, spherically symmetric black holes in General Relativity. The models considered are anisotropic perfect fluid dark matter (PFDM), isotropic constant-$\omega$ dark matter, and Bose Einstein condensate (BEC) dark matter with a polytropic equation of state (EoS). For each model, we derive the Einstein field equations and solve for the metric function analytically in the PFDM and constant-$\omega$ cases, and numerically via the Tolman-Oppenheimer-Volkoff equations for BEC dark matter. We then compute the key strong gravity observables: event horizon radius, photon sphere radius, black hole shadow radius, innermost stable circular orbit (ISCO), and circular orbital velocity profiles. The shadow radii obtained for each model are compared against Event Horizon Telescope constraints to place bounds on the dark matter parameters. Physical viability is assessed through the null, weak, dominant, and strong energy conditions, along with causality requirements on the sound speed. Our results show that PFDM produces near Schwarzschild geometry with mild anisotropic corrections, the constant-$\omega$ model is strongly restricted by causality to $0 \leq \omega \leq 1$, and BEC dark matter generates smooth, bounded deviations governed by the bosonic self interaction parameter $K$. Taken together, these findings show that strong gravity observables can serve as practical tools for distinguishing between competing dark matter models through their geometric imprints on black hole space times.

}

\begin{document}

\maketitle

\section*{ 1 Introduction}

One of the most important unsolved issues in contemporary physics is dark matter. The existence of a non-luminous matter component that dominates the mass budget of galaxies and clusters is strongly supported by observational data from gravitational lensing, galaxy rotation curves, and large-scale structure development \cite{rubin1970rotation,corbelli2000extended, clowe2006direct, collaboration2014planck}. Notwithstanding its significant gravitational impact, the fundamental characteristics of dark matter remain elusive. The mainstream cosmological model regards dark matter as cold and pressureless (Cold Dark Matter), although other possibilities incorporating pressure, self-interactions, or quantum condensate behavior have been extensively investigated \cite{kesden2006galilean, matos2000spherical, chavanis2011bec, hui2017ultralight}.\\
Compact objects and intense gravitational fields offer a complementary domain to investigate dark matter beyond cosmic findings. Specifically, alterations in spacetime geometry caused by dark matter halos can affect black hole horizons, photon spheres, accretion disk configuration, and orbital dynamics \cite{shaymatov2021effect, narzilloev2020dynamics, hou2018rotating, collaboration2019first}. Geometric signatures are becoming increasingly significant in the context of horizon-scale investigations conducted by the Event Horizon Telescope \cite{marongwe2023horizon, johnson2024black} and the precise measurement of accretion events. Thus, examining the alterations that various dark matter models impose on relativistic geometry provides a straightforward means to link dark matter microphysics with observable strong-gravity phenomena.\\
Numerous methodologies have modelled dark matter as an effective fluid source in the context of General Relativity. Anisotropic perfect fluid dark matter (PFDM) models implement regulated departures from Schwarzschild spacetime and can produce flattened galactic rotation curves via geometric modifications \cite{kiselev2003quintessence, salgado2003simple, kuncewicz2025impacts}. Isotropic fluid models characterized by a constant equation of state parameter $\omega$ yield Kiselev type solutions that integrate pressure influences into black hole geometry \cite{kiselev2003quintessence}. In the meantime, polytropic equations of state and smooth density profiles that are in line with halo phenomenology are produced by Bose Einstein condensate (BEC) dark matter models, which are inspired by ultra-light bosonic particles with self interactions \cite{chavanis2011bec, harko2011cosmological, boehmer2007can, chavanis2012growth}.\\
Notwithstanding the variety of these models, a comprehensive relativistic analysis of the impact of various dark matter equations of state on horizon structure, photon spheres, innermost stable circular orbits (ISCO), and circular velocities within a cohesive geometric framework is still constrained. Moreover, the prerequisites for physical viability namely the null, weak, dominant, and strong energy conditions, together with causality limits on sound speed are not uniformly examined across many models. These limitations are crucial for identifying which dark matter configurations are physically permissible in strong gravitational environments.\\
This study examines three typical equations of state for dark matter within a unified static, spherically symmetric framework: (i) anisotropic perfect fluid dark matter, (ii) isotropic constant-$\omega$ dark matter, and (iii) Bose Einstein condensate dark matter. For each scenario, we derive the associated Einstein field equations and ascertain the metric function either analytically or numerically via the Tolman-Oppenheimer-Volkoff framework. Subsequently, we calculate essential strong gravity observables, including the event horizon radius, photon sphere, ISCO radius, and circular orbital velocity profiles. Energy conditions and causality stipulations are established to obtain quantitative limitations on the model parameters. Our research elucidates how the features of dark matter manifest as visible geometric signatures by directly comparing anisotropic phenomenological, isotropic pressure supported, and microphysically justified dark matter models within a unified relativistic framework. This cohesive methodology establishes a framework for limiting dark matter microphysics via strong gravity phenomena.\\
The paper is organised as follows. In Section 2 we describe the theoretical background including the static spherically symmetric metric, the stress-energy tensor and the corresponding Einstein field equations. In section 3 we present the three equations of state for dark matter considered in this study anisotropic perfect fluid dark matter, constant-$\omega$ dark matter and Bose Einstein condensate dark matter and obtain the corresponding field equations. In Section 4 we calculate the main strong gravity observables, i.e. the event horizon radius, photon sphere, innermost stable circular orbit (ISCO), and circular orbital velocity profiles. Section 5 is dedicated to the analysis of energy requirements and causality restrictions for each model. In section 6 we present numerical constraints on the model parameters and discuss their physical implications. Section 7 concludes our findings and suggests possible directions for future research.

\section*{2 Theoretical Framework}
A compact object immersed in a dark matter distribution is described by a static, spherically symmetric spacetime. The line element in Schwarzschild like coordinates $(t,r,\theta,\phi)$ is expressed as
\begin{equation}
    ds^2=-f(r)dt^2+\frac{dr^2}{f(r)}+r^2d\theta^2+r^2sin^2\theta d\phi^2
\end{equation}
where the metric function $f(r)$ completely defines the spacetime geometry. It is expedient to articulate $f(r)$ in relation to the mass function m(r) as
\begin{equation}
    f(r)=1-\frac{2m(r)}{r}
\end{equation}
This form ensures that the spacetime reduces to the Schwarzschild solution in the vacuum limit, $m(r) \rightarrow M=constant$.\\
Dark matter is represented as an effective relativistic fluid characterized by the stress-energy tensor \cite{letelier1979clouds, herrera1997local}
\begin{equation}
    T^\mu _\nu =diag(-\rho(r),p_r(r),p_t(r),p_t(r)),
\end{equation}
where $\rho(r)$ represents the energy density and $p_r(r)$ and $p_t(r)$ denote the radial and tangential pressures respectively. This form accommodates both isotropic $(p_r = p_t = p)$ and anisotropic dark matter configurations, the latter being strongly justified in relativistic halo and compact object contexts.\\
Einstein’s field equations (with G=c=1),
\begin{equation}
    G^\mu_\nu=8\pi T^\mu_\nu
\end{equation}
lead to the standard mass relation
\begin{equation}
    \frac{dm}{dr}=4\pi r^2 \rho(r)
\end{equation}
which characterizes the radial distribution of the enclosed gravitational mass. Conservation of the stress–energy tensor, $\Delta_{\mu} T_{\nu}^{\mu}=0$, yields  generalized Tolman-Oppenheimer-Volkoff equation,
\begin{equation}
    \frac{dp_r}{dr}=-\frac{(\rho+p_r)[m(r)+4\pi r^3p_r]}{r[r-2m(r)]}+\frac{2}{r}(p_t-p_r)
\end{equation}
which reduces to the standard isotropic form
\begin{equation}
    \frac{dp}{dr}=-\frac{(\rho +p)[m(r)+4\pi r^3p]}{r[r-2m(r)]}
\end{equation}
when $(p_r = p_t = p)$. Equations (5)–(7), in conjunction with a defined EoS linking pressure and density, fully ascertain the spacetime geometry.

\section*{3 Dark Matter Equation of State}
To examine the impact of dark matter microphysics on spacetime geometry, we analyze three sample equations of state (EoS): anisotropic perfect fluid dark matter (PFDM), isotropic constant-$\omega$ dark matter, and Bose Einstein condensate (BEC) dark matter. These models encompass phenomenological, analytic, and microphysically grounded representations of dark matter, facilitating a systematic comparison within a unified relativistic framework.\\

We initially examine anisotropic perfect fluid dark matter, which has been extensively utilized to represent relativistic dark matter halos encircling small objects \cite{kiselev2003quintessence, salgado2003simple, shaymatov2021effect}. In this approach, anisotropy is introduced through the relations
\begin{equation}
    p(r)=0 , \quad p(t)=(1-\epsilon)\rho
\end{equation}
where $\epsilon$ is a dimensionless parameter that governs deviations from vacuum geometry. Substituting Equation (8) into the Einstein equations results in a first-order differential equation for the metric function. The integration of this equation results in the following \cite{li2012galactic, xu2020black}
\begin{equation}
    f(r)=1-\frac{2M}{r}+\frac{r^{2(1-\epsilon)}}{r_e}
\end{equation}
where M is the central mass and $r_e$ represents a fixed length scale linked to the dispersion of dark matter. This metric reduces to the Schwarzschild solution when $r_e \rightarrow \infty$ and $\epsilon \rightarrow 1$.\\
Next, we examine isotropic dark matter characterised by a constant parameter in the equation of state,
\begin{equation}
    p=\omega \rho
\end{equation}
where $\omega$ is a constant. This form has been investigated in the context of Kiselev type black hole solutions and offers the most straightforward extension of pressureless cold dark matter \cite{kiselev2003quintessence}. Solving the field equations for this EoS yields
\begin{equation}
    f(r)=1-\frac{2M}{r}-\frac{C}{r^{3\omega+1}}
\end{equation}
where C is an integration constant associated with the amplitude of dark matter density. The supplementary power-law correction is expressly contingent upon $\omega$, with the Schwarzschild limit being reinstated as $C\rightarrow0$.\\
We ultimately examine Bose Einstein condensate (BEC) dark matter, inspired by ultralight bosonic particles exhibiting repulsive self-interactions \cite{boehmer2007can, harko2011cosmological, chavanis2011bec}.In the Thomas–Fermi approximation, the equation of state has a polytropic form
\begin{equation}
    p=K\rho^2
\end{equation}
where $K$ represents the microscopic characteristics of the bosonic particles. In this case, the field Equations (5)–(7) necessitate numerical integration. The spacetime metric preserves its general structure
\begin{equation}
    f(r)=1-\frac{2m(r)}{r}
\end{equation}
where the mass function m(r) is obtained by solving
\begin{equation}
    \frac{dm}{dr}=4\pi r^2\rho(r), \quad \frac{d\rho}{dr}= -\frac{(1+K\rho)[m(r)+4\pi r^3K\rho^2]}{2Kr(r-2m(r))}
\end{equation}
Equations (9), (11), and (13) collectively delineate the three spacetime geometries examined in this study. We systematically evaluate how dark matter microphysics is represented in strong gravity observables by comparing their horizon structure, photon spheres, orbital dynamics, and energy-condition restrictions.

\section*{4 Strong-Gravity Observables for Dark Matter Models}
This section calculates the primary strong gravity observables for each dark matter equation of state presented in Section 3.
\subsection*{ 4.1 Perfect Fluid Dark Matter}
We initially examine the anisotropic perfect fluid dark matter (PFDM) model, which characterizes dark matter as an effective anisotropic fluid that alters the Schwarzschild geometry. This methodology quantifies the gravitational impact of dark matter halos via a controlled deviation parameter $\epsilon$, enabling the examination of how anisotropy modifies strong-gravity observables.\\\\
\textbf{Event Horizon}\\\\
The event horizon is located at the radius $r_{h}$ at which the temporal metric component becomes null, i.e.,
\begin{equation}
    e^{\nu(r_h)}=0
\end{equation}
Since $e^{\nu(r)}=f(r)$, this condition becomes $f(r_h)=0.$\\
By substituting the PFDM metric function presented in Section 3, we obtain
\begin{equation}
    1-\frac{2M}{r_h}+\frac{r_h^{2(1-\epsilon)}}{r_e}=0
\end{equation}
Resolving this equation produces the horizon radius $r_h$. As $\epsilon$ approaches 1, the Schwarzschild value $r_h=2M$ is obtained. \\\\
\textbf{Photon Sphere}\\\\
The photon sphere pertains to unstable circular null geodesics and is defined by the condition
\begin{equation}
    \frac{d}{dr}\left(\frac{e^{\nu(r)}}{r^2}\right)=0
\end{equation}
This is equivalent to
\begin{equation}
    2f(r_{ph})=r_{ph}f'(r_{ph})
\end{equation}
Substituting the PFDM metric function yields
\begin{equation}
    2-\frac{6M}{r_{ph}}+\frac{2\epsilon}{r_e}r_{ph}^{2(1-\epsilon)}=0
\end{equation}
\textbf{Circular Orbital Velocity}\\\\
The tangential velocity linked to a circular orbit is derived from the conventional geodesic analysis of static, spherically symmetric spacetimes \cite{kuncewicz2025perfect}, and is expressed as follows
\begin{equation}
    v^2(r)=\frac{rf'(r)}{2}
\end{equation}
Substituting the PFDM metric function yields
\begin{equation}
    v^2(r)=\frac{M}{r}+\frac{(1-\epsilon)}{r_e}r^{2(1-\epsilon)}
\end{equation}
\textbf{Innermost Stable Circular Orbit (ISCO)}\\\\
The ISCO represents the minimum of the effective potential for massive test particles. It is determined by resolving
\begin{equation}
    \partial _r V_{eff}(r_{ISCO})=0, \qquad \partial_r^2V_{eff}(r_{ISCO})=0
\end{equation}
Using the effective potential
\begin{equation}
    V_{eff}(r)=f(r)\left(1+\frac{L^2}{r^2}\right)
\end{equation}
the initial condition produces the angular momentum of circular orbits,
\begin{equation}
    L^2(r)=\frac{r^3f'(r)}{2f(r)-rf'(r)}
\end{equation}
Enforcing the marginal stability requirement $\partial_rL^2(r)=0$, equivalently $\partial_r^2V_{eff}(r)=0$, results in the overarching ISCO equation \cite{bardeen1972rotating, pugliese2011motion},
\begin{equation}
    3f(r)f'(r)-2r(f'(r))^2+rf(r)f''(r)=0
\end{equation}
By using the PFDM metric function presented in Section 3, equation (26) becomes,
\begin{equation}
    2\frac{( -6M^2+\frac{2r^{4(1-\epsilon)}(\epsilon-1)(r^{2\epsilon}r_e(\epsilon-2)-r^2\epsilon)}{r_e^2}+\frac{Mr(r_e+r^{2(1-\epsilon)}(-15+20\epsilon-4\epsilon^2))}{r_e}}{r^3}=0
\end{equation}
This algebraic equation calculates the ISCO radius $r_{ISCO}$ for the PFDM spacetime. As $\epsilon \rightarrow 1$, the Schwarzschild value $r_{ISCO}=6M$ is obtained.\\\\
\subsection*{4.2 Constant -$\omega$ Dark Matter}
We subsequently examine constant-$\omega$ dark matter, represented as an isotropic fluid characterized by an equation of state $p=\omega \rho$. The parameter $\omega$ regulates pressure effects and dictates the deviations of the horizon, photon sphere, circular velocity, and ISCO from their Schwarzschild values.\\\\
\textbf{Event Horizon}\\\\
Inserting the metric function into the horizon condition produces
\begin{equation}
    1-\frac{2M}{r_h}-\frac{C}{r_h^{3\omega+1}}=0
\end{equation}
\begin{equation}
    \Longrightarrow \qquad r_h^{3\omega+1}-2Mr_h^{3\omega}-C=0
\end{equation}
\\
This equation identifies the position of the event horizon in the context of pressure-supported dark matter. The supplementary term proportional to $C$ alters the Schwarzschild radius based on the sign and magnitude of $\omega$.\\\\
\textbf{Photon sphere}\\\\
By applying the photon-sphere condition, we obtain
\begin{equation}
    2-\frac{6M}{r_{ph}}-3(\omega +1)\frac{C}{r_{ph}^{3\omega+1}}=0
\end{equation}
The inclusion of the dark matter term alters the radius of unstable circular photon orbits. The reliance on $3\omega+1$ indicates that pressure directly affects light propagation in proximity to the compact object.\\\\
\textbf{Circular Orbital Velocity}\\\\
Inserting the metric function into the velocity equation yields
\begin{equation}
    v^2(r)=\frac{M}{r}+\frac{(3\omega+1)C}{2r^{3\omega +1}}
\end{equation}
The second term denotes the adjustment attributable to pressure supported dark matter. The radial behavior is determined by $3\omega+1$, signifying that increased values of $\omega$ amplify deviations from the Keplerian falloff.\\\\
\textbf{Innermost Stable Circular Orbit (ISCO)}\\\\
Inserting the constant-$\omega$ metric function inside the previously established ISCO equation (26), we derive
\begin{equation}
    r^{-3-6\omega}[-2M(6M-r)r^{6\omega}-3C^2(1+4\omega+3\omega^2)+Cr^{3\omega}(r-9r\omega^2
    +6M(-2-4\omega+3\omega^2))]=0
\end{equation}
The solutions to this equation ascertain the radius of the innermost stable circular orbit, $r_{ISCO}$, for the constant-$\omega$ dark matter configuration. The inclusion of the parameter $\omega$ alters the stability characteristics of circular orbits, resulting in adjustments to the $ISCO$ radius compared to the vacuum scenario.\\\\
\subsection*{4.3 Bose Einstein Condensate (BEC) Dark Matter}
We now examine Bose Einstein Condensate dark matter, described by the polytropic equation of state $p=K\rho^2$. The self interaction parameter $K$ governs the density profile and its impact on the horizon, photon sphere, circular velocity and ISCO structure.\\\\
\textbf{Event Horizon}\\\\
Incorporating the metric function into the horizon condition yields
\begin{equation}
    1-\frac{2m(r_h)}{r_h}=0
\end{equation}
The horizon radius is contingent upon the cumulative mass distribution as defined by the TOV equations. In contrast to the analytic models, the horizon structure is determined by the self-interaction parameter $K$ and the central density. The Schwarzschild limit is attained when the contribution of dark matter is rendered insignificant.\\\\
\textbf{Photon sphere}\\\\
Utilizing the photon-sphere condition, we derive
\begin{equation}
    r\left(\frac{2m(r)}{r^2}-\frac{2m'(r)}{r}\right)=2\left(1-\frac{2m(r)}{r}\right)
\end{equation}
\begin{equation}
    \Longrightarrow \qquad  m'(r)=\frac{3m(r_{ph})}{r_{ph}}-1
\end{equation}
The photon sphere is explicitly contingent upon both the enclosed mass and its radial gradient, indicating the uniform matter distribution of the condensate. The alteration concerning Schwarzschild is thus governed by the density profile derived from numerical integration.\\\\
\textbf{Circular Orbital Velocity}\\\\
When the velocity equation is amended with the metric function, then
\begin{equation}
    v^2=\frac{m(r)}{r}-m'(r)
\end{equation}
where $m'(r)=4\pi r^2 \rho(r)$. The velocity profile is directly influenced by the mass function and its derivative, resulting in smooth, constrained deviations from Keplerian behavior. The self interaction strength $K$ governs the magnitude of these alterations via its effect on $\rho(r)$.\\\\
\textbf{ISCO radius}\\\\
Inserting the metric function and its derivatives in equation (26),
\begin{equation}
    r\left(1-\frac{2m}{r}\right) \left(\frac{-2m''}{r}+\frac{4m'}{r^2}-\frac{4m}{r^3}\right)-2r\left(\frac{-2m'}{r}+\frac{2m}{r^2}\right)^2+3\left(1-\frac{2m}{r}\right)\left(\frac{-2m'}{r}+\frac{2m}{r^2}\right)=0
\end{equation}
The radius of the ISCO depends on the specific shape of the effective potential, which in turn depends on the numerical metric. Fluctuations in central density and the parameter $K$ modify the curvature of the potential, hence altering the initiation of orbital instability.\\\\
\section*{Energy Conditions and Causality}
Energy conditions offer essential criteria for assessing the physical admissibility of a specific matter configuration in General Relativity \cite{visser2018lorentz, kontou2020energy}. They are extensively utilized to investigate relativistic fluids, compact objects, and altered spacetime geometries \cite{hawking2023large, wald2000general, poisson2004relativist, carroll2004introduction}. Regarding the anisotropic stress-energy tensor presented in Section 2,
\begin{equation}
    T^\mu _\nu=diag(-\rho, p_r, p_t, p_t)
\end{equation}
the standard energy conditions can be articulated as inequalities that encompass the energy density and pressure components.\\
The \textbf{null energy condition (NEC)} requires
\begin{equation}
    \rho+p_i\geq 0    \qquad  (i=r,t)
\end{equation}
while the \textbf{weak energy condition (WEC)} demands
\begin{equation}
        \rho \geq 0, \qquad  \rho+p_i \geq 0
\end{equation}
The \textbf{dominant energy condition (DEC)} imposes
\begin{equation}
    \rho \geq |p_i|
\end{equation}
ensuring that the energy flux observed by any observer remains causal \cite{hawking2023large, wald2000general}. Finally, the \textbf{strong energy condition (SEC)} is written as
\begin{equation}
    \rho +p_r+2p_t \geq 0
\end{equation}
which plays an important role in gravitational collapse and cosmological dynamics \cite{poisson2004relativist}.\\
Moreover, relativistic fluids must adhere to a causality constraint that limits the propagation speed of minor perturbations to the sound speed
\begin{equation}
    c_s^2=\frac{dp}{d\rho}
\end{equation}
with the physical conditon $0\leq c_s^2 \leq 1$.
This stipulation guarantees that disturbances propagate at velocities not beyond the speed of light \cite{poisson2004relativist, carroll2004introduction}.\\
In the PFDM model presented in Section 3, the pressures comply with
\begin{equation}
    p_r=0, \qquad p_t=(1-\epsilon)\rho
\end{equation}
Inserting these relations into the aforementioned conditions produces
\begin{equation}
    \rho+p_r=\rho \geq 0
\end{equation}
and 
\begin{equation}
    \rho+p_t=\rho(2-\epsilon) \geq 0
\end{equation}
Therefore the NEC and WEC are satisfied provided $0 \leq \epsilon \leq 2$.\\
The dominant energy condition requires $\rho \geq |p_t|$, which results in the anisotropy parameter having the same bound. The strong energy condition becomes
\begin{equation}
    \rho+p_r+2p_t= \rho(3-2\epsilon) \geq 0,
\end{equation}
which holds for $\epsilon \leq \frac{3}{2}$. Therefore, although permitting deviations from the Schwarzschild geometry, PFDM configurations with moderate anisotropy satisfy the typical relativistic energy criteria.\\\\
For the constant-$\omega$ dark matter model characterized by its equation of state $p=\omega \rho$, the NEC and WEC require
\begin{equation}
    \rho + p = \rho (1+\omega) \geq 0  
\end{equation}
implying $\omega \geq -1$.\\
The dominant energy condition imposes $\omega \leq 1$, while the strong energy condition gives
\begin{equation}
    \rho+3p= \rho(1+3 \omega) \geq 0
\end{equation}
which holds for $\omega \geq - \frac{1}{3}$. The sound speed for this model is $c_s^2= \omega$, and causality therefore requires $0 \leq \omega \leq 1$.\\\\
Ultimately, for the Bose–Einstein condensate dark matter hypothesis characterized by its equation of state $p=K \rho^2$, the NEC and WEC are inherently fulfilled as
\begin{equation}
    \rho + p = \rho (1+K \rho) \geq 0
\end{equation}
The dominant energy condition requires $\rho \geq p$, which leads to the constraint $ \rho \leq \frac{1}{K}$.\\
The strong energy condition becomes
\begin{equation}
    \rho+3p = \rho(1+3K\rho) \geq 0
\end{equation}
which is always satisfied for positive density. The sound speed is $c_s^2 = 2K\rho$, and causality therefore imposes $\rho \leq \frac{1}{2K}$ \cite{colpi1986boson}.\\
The results indicate that all three dark matter models fulfill the fundamental physical criteria of relativistic fluids within suitable parameter ranges. Specifically, PFDM facilitates regulated anisotropic modifications to vacuum geometry, constant-$\omega$ dark matter is significantly restricted by causality, and BEC dark matter inherently complies with energy conditions while imposing density limitations via the sound-speed constraint.
    
\section*{Discussion}

To comprehend how various dark matter equations of state alter the spacetime geometry, we initially examine the behavior of the metric function f(r).

\begin{figure}[H]
    \centering
    \includegraphics[width=1\textwidth]{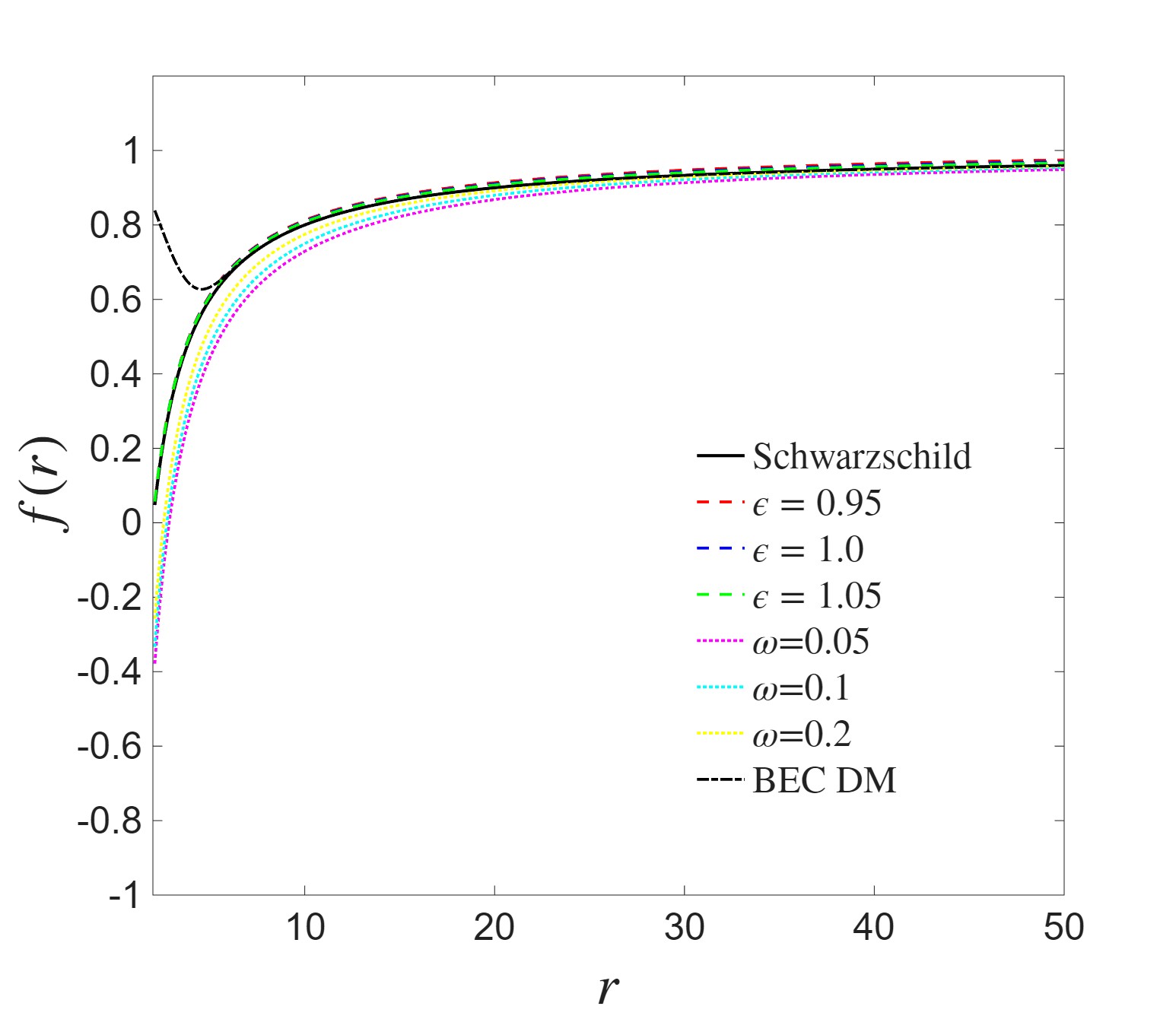}
    \caption{Radial variation of the metric function $f(r)$ across various dark matter theories in comparison to the Schwarzschild spacetime. The curves demonstrate the alterations in spacetime geometry induced by PFDM ($\epsilon$), constant-$\omega$ dark matter, and BEC dark matter, while maintaining asymptotic flatness at extensive distances. }
    
\end{figure}

Figure 1 depicts the radial characteristics  $f(r)$ for several dark matter EoS, alongside the Schwarzschild solution utilized as a reference spacetime. The Schwarzschild metric increases continuously with radial distance and approaches $f(r) \rightarrow1$ at extensive radii, signifying the restoration of asymptotically flat spacetime.\\
These results indicate that various dark matter equations of state provide unique signatures on the spacetime geometry around compact objects. Anisotropic dark matter (PFDM) yields regulated geometric corrections, isotropic pressure in the constant-$\omega$ model alters the gravitational potential, and the BEC model induces smooth spacetime deformations dictated by the underlying density profile. These alterations are anticipated to affect measurable strong-gravity parameters, including photon spheres, accretion disk behavior, and orbital dynamics.\\
Figure 2 depicts the behavior of the metric function f(r) for the PFDM model across various values of the anisotropy parameter $\epsilon$. The event horizon is defined as the radial position where $f(r)=0$, represented by the intersection with the horizontal dashed line. The plot indicates that the horizon is located at $r/M \approx 2$, akin to the Schwarzschild scenario; nevertheless, minor adjustments in $\epsilon$ result in modest alterations in the metric function near horizon.\\
\begin{figure}[H]
    \centering
    \includegraphics[width=1\textwidth]{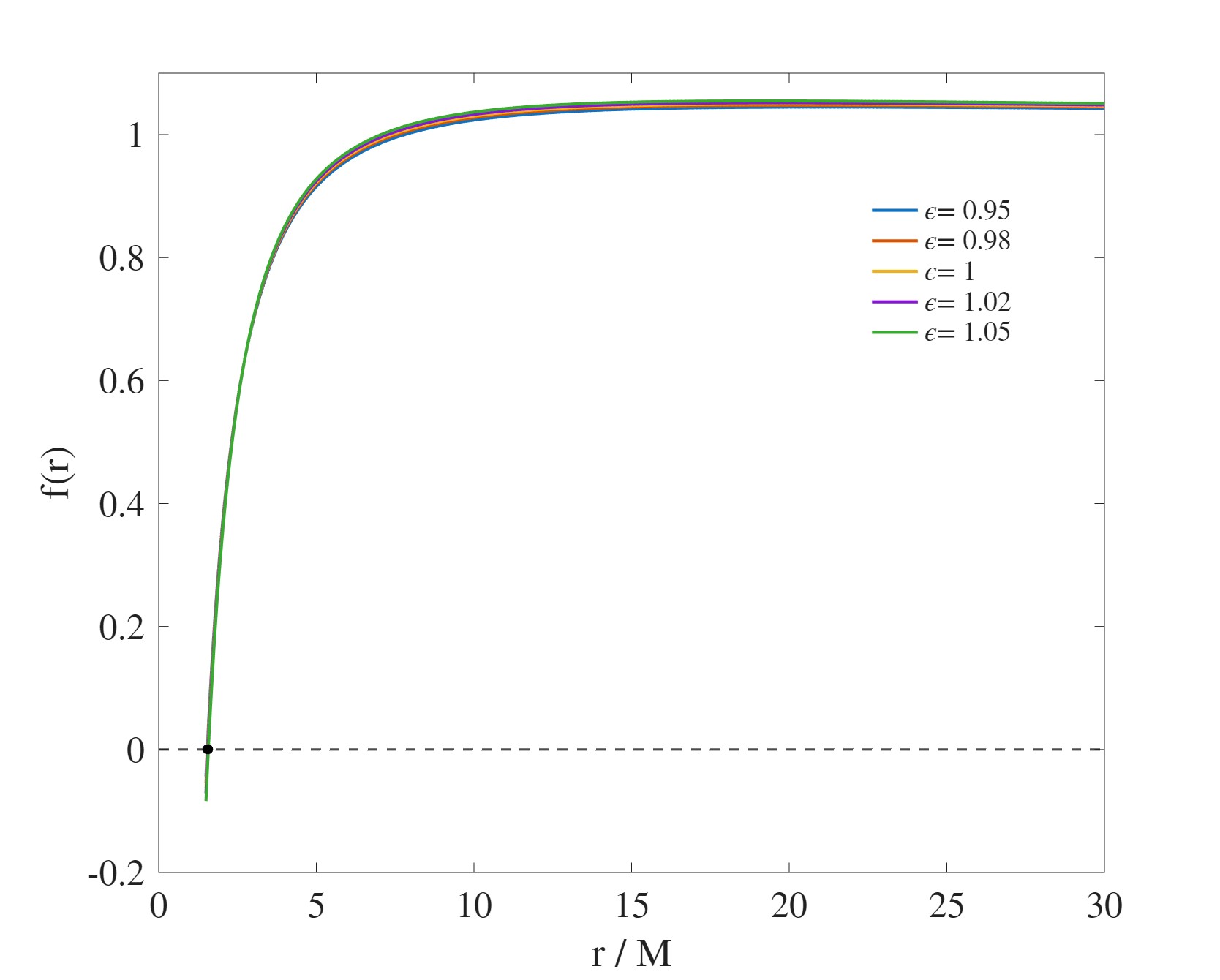}
    \caption{The metric function f(r) for the PFDM spacetime illustrates the event horizon location for distinct values of the anisotropy coefficient $\epsilon$. The event horizon is defined by the intersection f(r)=0, indicated by dashed horizontal line. }
    
\end{figure}
This suggests that anisotropic dark matter alters the spacetime geometry around the compact object while maintaining the event horizon's presence. The parameter $\epsilon$ regulates the intensity of the dark matter contribution, with more deviations from $\epsilon=1$ resulting in somewhat enhanced alterations in the gravitational potential. Nevertheless, the horizon stays unchanged, indicating that PFDM functions as a perturbative modification to the Schwarzschild spacetime rather than fundamentally transforming the black hole's structure.\\
Figure 3 illustrates the behavior of the metric function f(r) for various values of the EoS parameter $\omega$ in the constant-$\omega$ dark matter scenario. The event horizon is defined as the radial position where f(r)=0, shown by the dashed horizontal line. The solution for $\omega=0$ exhibits features resembling the usual Schwarzschild spacetime. As the value of $\omega$ fluctuates, the metric function is changed. This causes changes in the horizon structure . In particular, positive values of $\omega$ slightly improve the metric function, while negative values could significantly change the geometry and possibly change the position of the horizon..
\begin{figure}[H]
    \centering
    \includegraphics[width=1\textwidth]{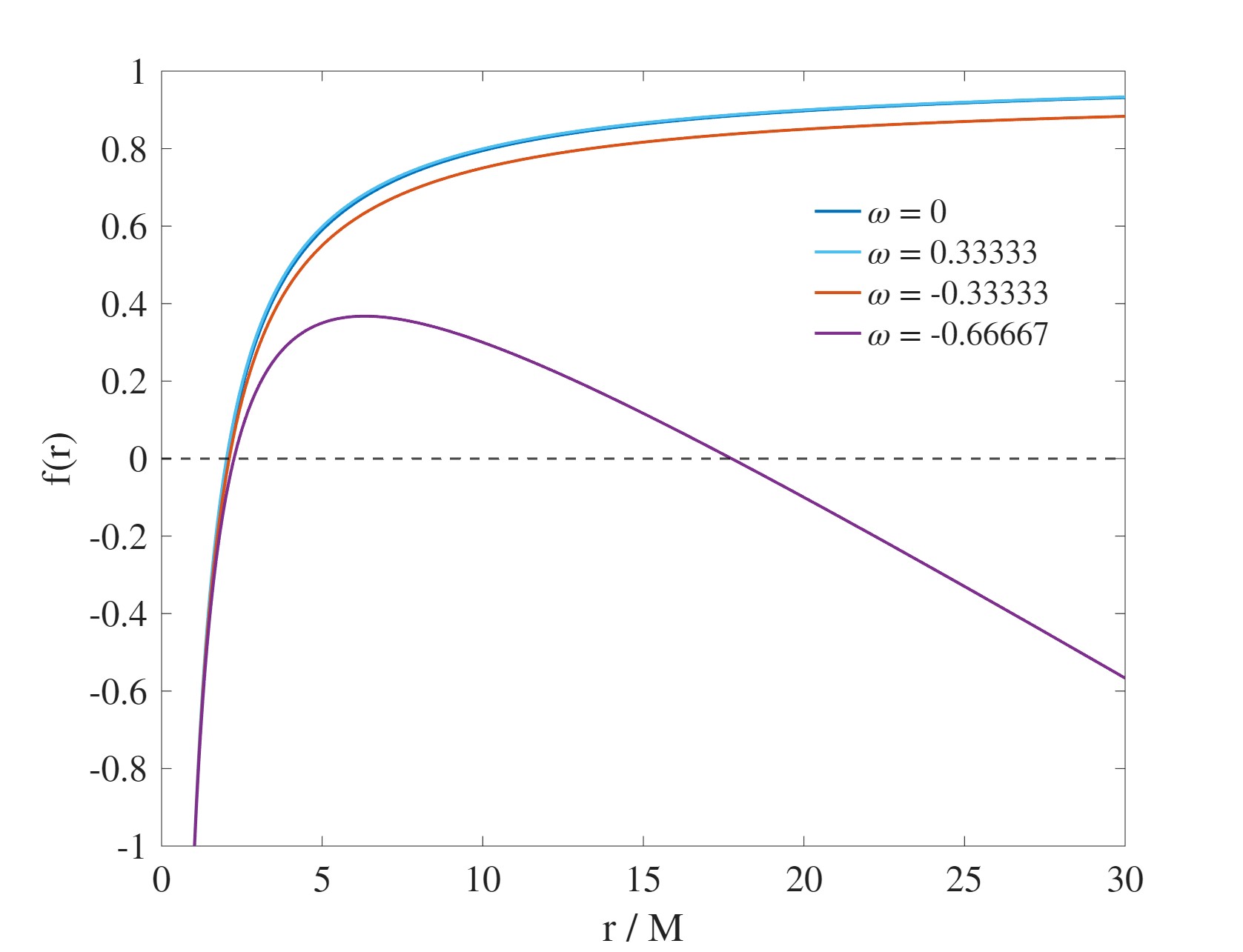}
    \caption{Metric function $f(r)$ for the constant $\omega$ dark matter model for different values of the EoS parameter $\omega$. }
\end{figure}

This trend shows concretely the effect of the isotropic dark matter pressure on the curvature of spacetime. The parameter $\omega$ controls the correlation between pressure and energy density and its variation changes the gravitational potential around the compact object. The large negative values of the EoS parameter can significantly modify the spacetime geometry and can hinder the formation of a stable horizon, indicating the importance of the equation of state parameter in the physical viability of constant-omega dark matter configurations.\\

\begin{figure}[H]
    \centering

    \begin{subfigure}{0.32\textwidth}
        \centering
        \includegraphics[width=\linewidth]{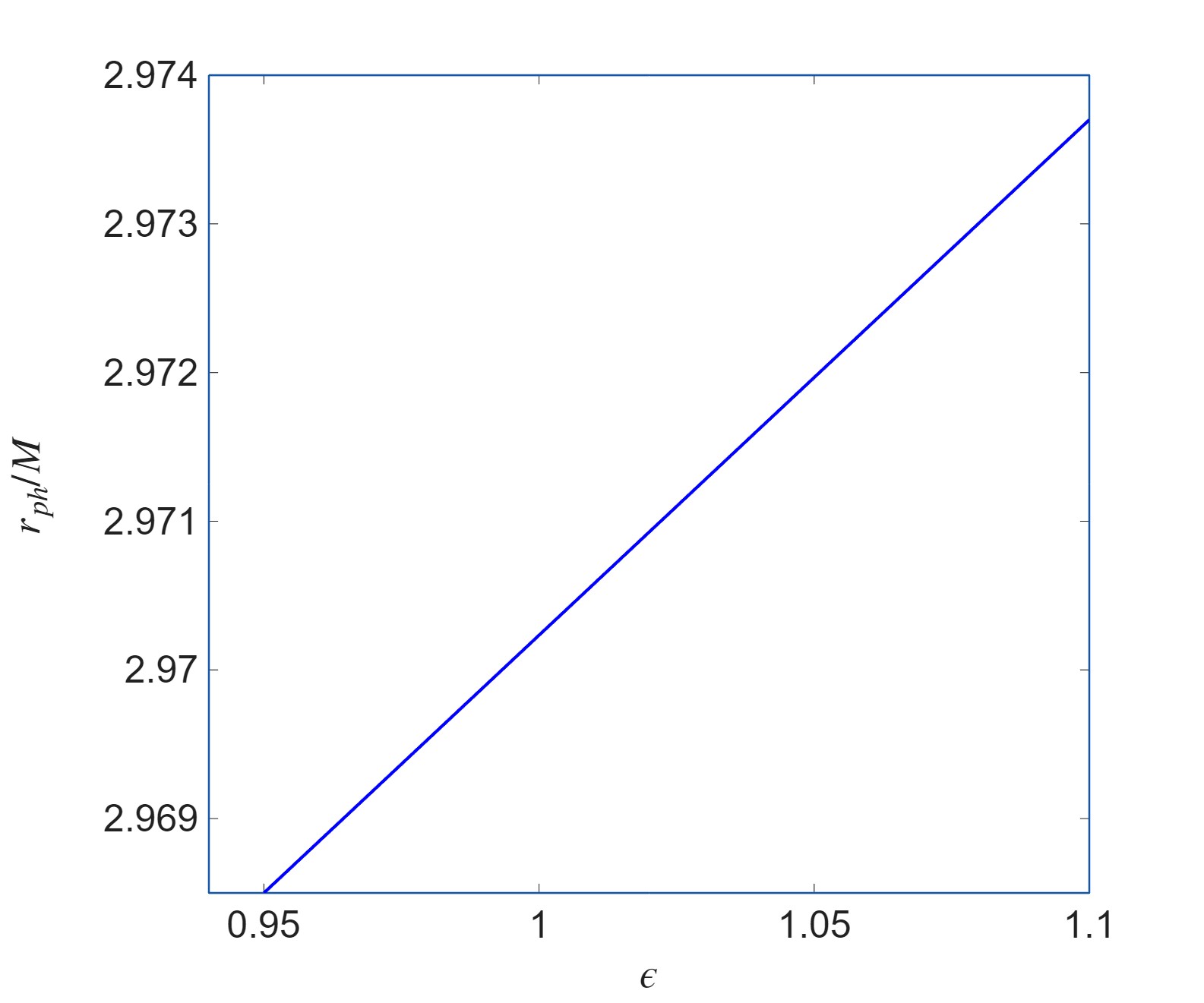}
        \caption{}
    \end{subfigure}
    \hfill
    \begin{subfigure}{0.32\textwidth}
        \centering
        \includegraphics[width=\linewidth]{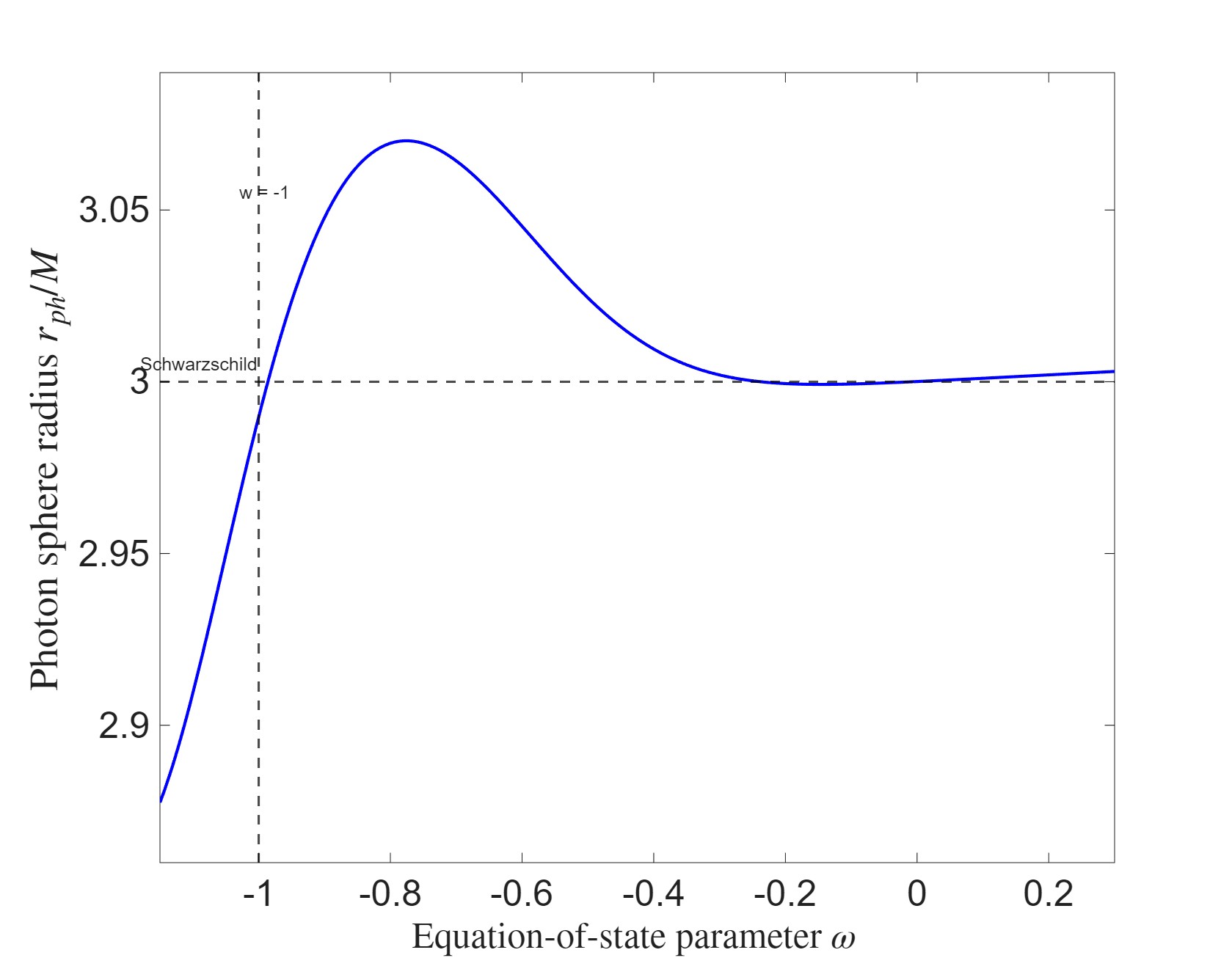}
        \caption{}
    \end{subfigure}
    \hfill
    \begin{subfigure}{0.32\textwidth}
        \centering
        \includegraphics[width=\linewidth]{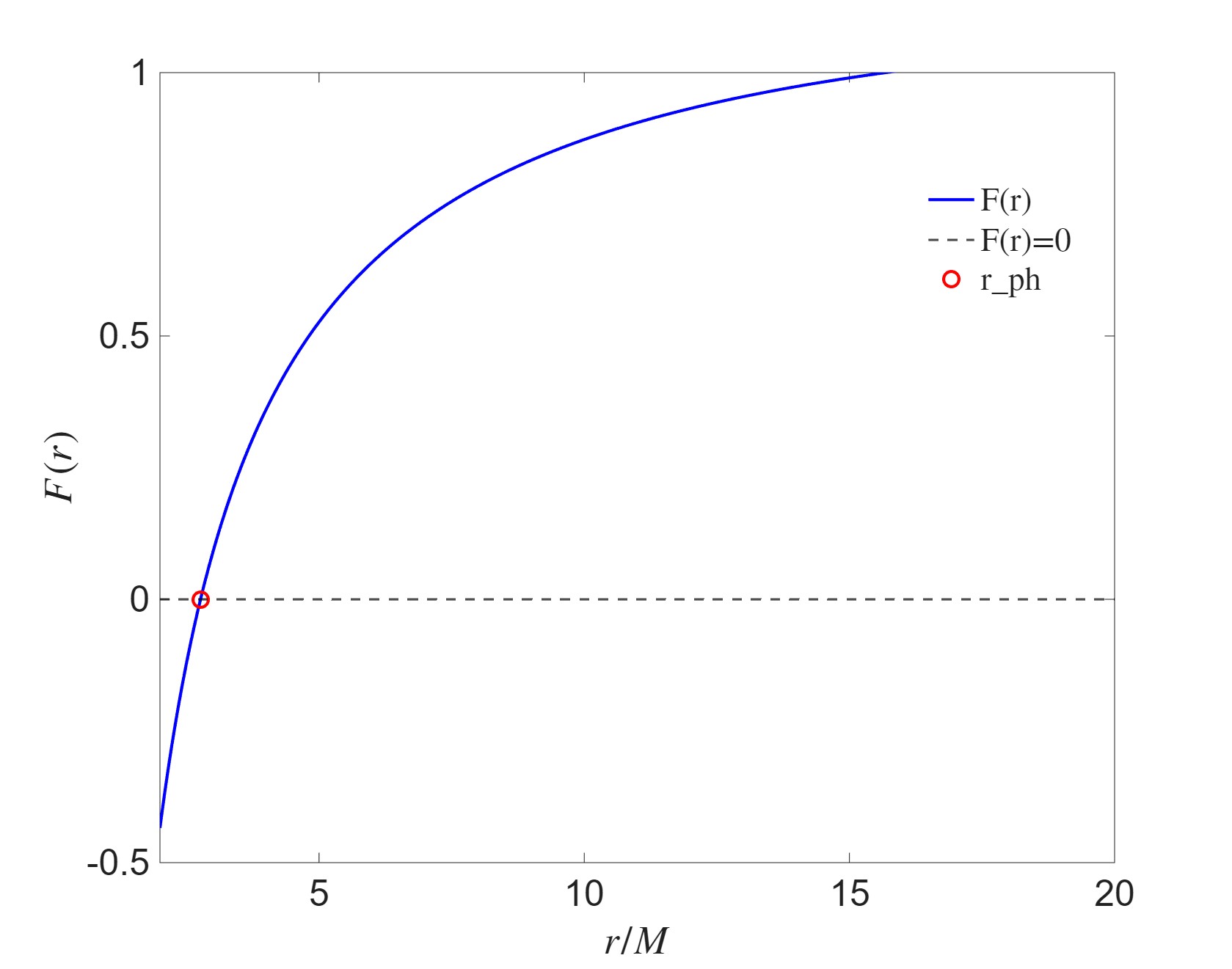}
        \caption{}
    \end{subfigure}

    \caption{Fig. 4: Behavior of the photon sphere for different dark matter models. (a) PFDM: variation of $r_{ph}/M$ with $\epsilon$. (b) Constant-$\omega$ model showing dependence on $\omega$. (c) BEC model where Photon sphere is achieved from $f(r) = 0$.}

\end{figure}
Figure 4 illustrates that $r_{ph}$ is influenced by the dark matter equation of state, establishing a clear connection between dark matter microphysics and observable strong-gravity phenomena. In the PFDM model, the photon sphere demonstrates minimal departures from the Schwarzschild value, suggesting that anisotropic dark matter functions as a perturbative correction to vacuum spacetime. Conversely, the constant-$\omega$ model exhibits a heightened reliance on the EoS parameter, especially for negative $\omega$, where substantial variations occur due to the effective pressure's influence on the gravitational field.\\
These results indicate that strong gravity observables, such as the photon sphere, which are closely associated with black hole shadow dimensions and gravitational lensing, may function as possible probes of dark matter characteristics. This creates a link between cosmological dark matter models and observable relativistic phenomena around compact objects.
\begin{figure}[H]
    \centering
    \includegraphics[width=1\textwidth]{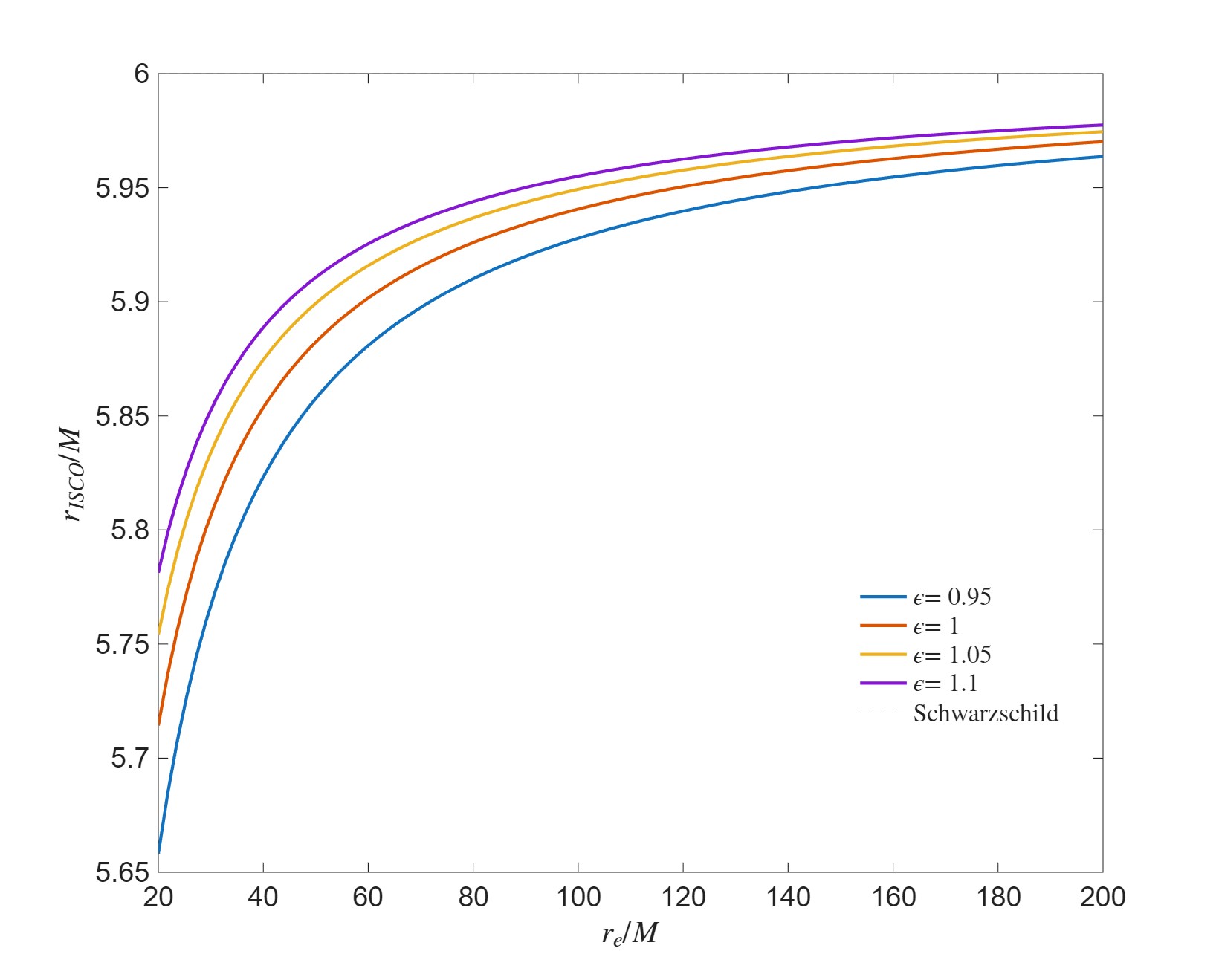}
    \caption{For varying values of the PFDM parameter $\epsilon$, the ISCO radius $r_{ISCO}/M$ varies as a function of the dark matter scale parameter $r_e/M$. }
\end{figure}
The diagram demonstrates how perfect fluid dark matter (PFDM) alters the position of the innermost stable circular orbit. For diminished values of the dark matter scale $r_e$, the ISCO radius markedly diverges from the Schwarzschild value, signifying a more pronounced impact of dark matter on the near-horizon geometry. As $r_e$ rises, all curves asymptotically converge to the Schwarzschild limit $r_{ISCO}=6M$, indicating that the influence of dark matter diminishes at extensive scales. Moreover, the parameter $\epsilon$ governs the magnitude of this deviation: increased values of $\epsilon$ result in a more pronounced inward displacement of the ISCO radius, indicating more substantial alterations to the gravitational potential. This pattern indicates that PFDM can modify the inner edge of accretion disks surrounding compact objects, potentially resulting in detectable traces in high energy astrophysical events such as black hole accretion and emission spectra.
\begin{figure}[H]
    \centering
    \includegraphics[width=1\textwidth]{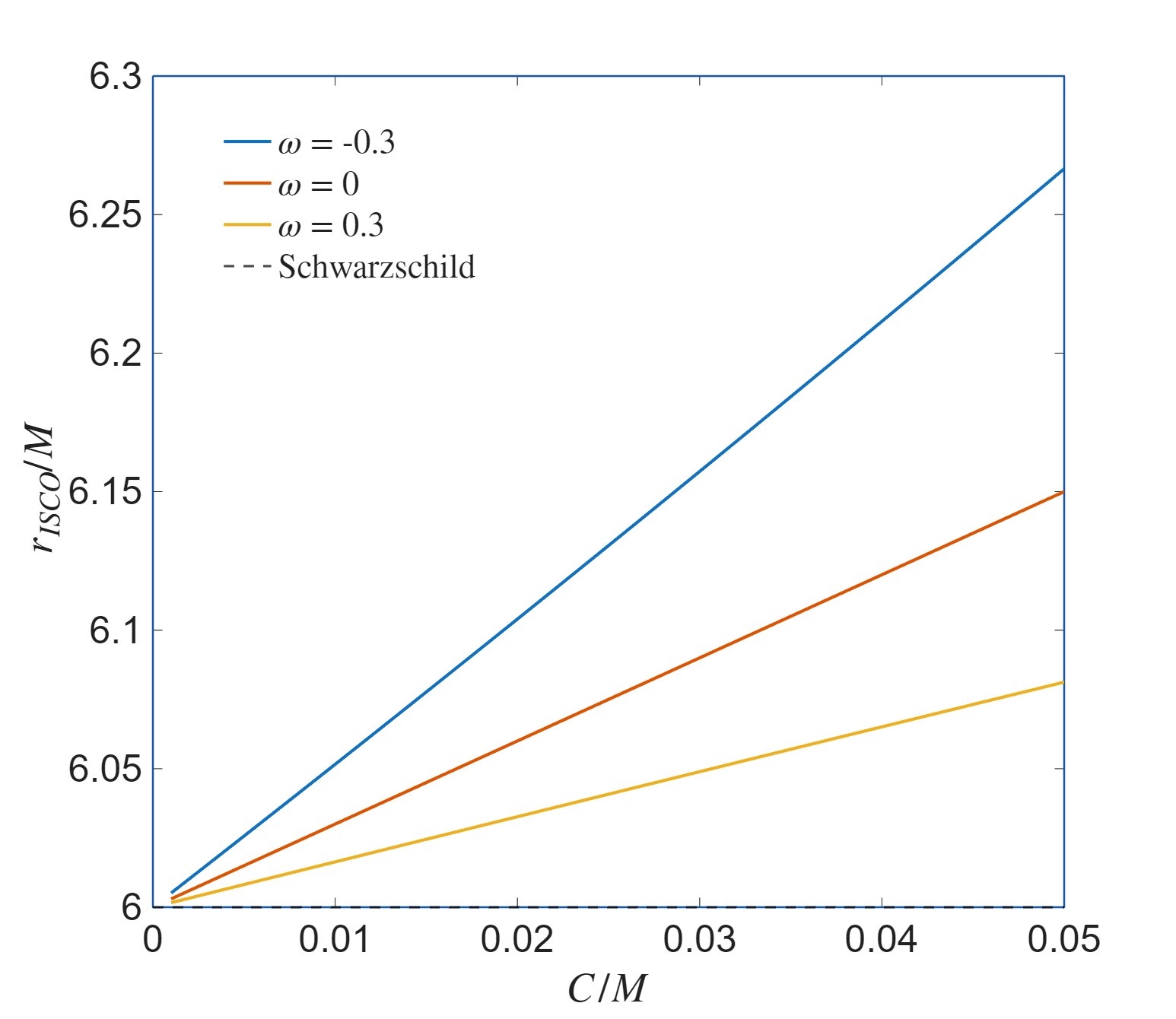}
    \caption{Variations of the ISCO radius $r_{ISCO}/M$ with respect to the dark matter strength parameter $C/M$ for various values of $\omega$. }
\end{figure}
The graph shows that the ISCO radius increases monotonically with the dark matter strength parameter $C$, which means the presence of dark matter pushes the inner edge of the accretion disk outwards compared to the Schwarzschild situation. The external displacement is more and more clear when the equation of state parameter $\omega$ decreases. In particular, the negative values of $\omega$ (e.g. $\omega=-0.3$) give the maximum departure, indicating a more enhanced effective gravitational modification due to dark matter. On the other hand, positive values of $\omega$ lead to smaller deviations, suggesting less effect on the geometry of the spacetime. We find that all curves tend to the Schwarzschild limit $r_{ISCO} = 6M$ as C approaches 0, confirming the consistency of the model. The results show that the density and pressure characteristics of dark matter, represented by C and $\omega$, can significantly affect the structure of the accretion disks and provide detectable imprints in black hole astrophysics.
\begin{figure}[H]
    \centering
    \includegraphics[width=1\textwidth]{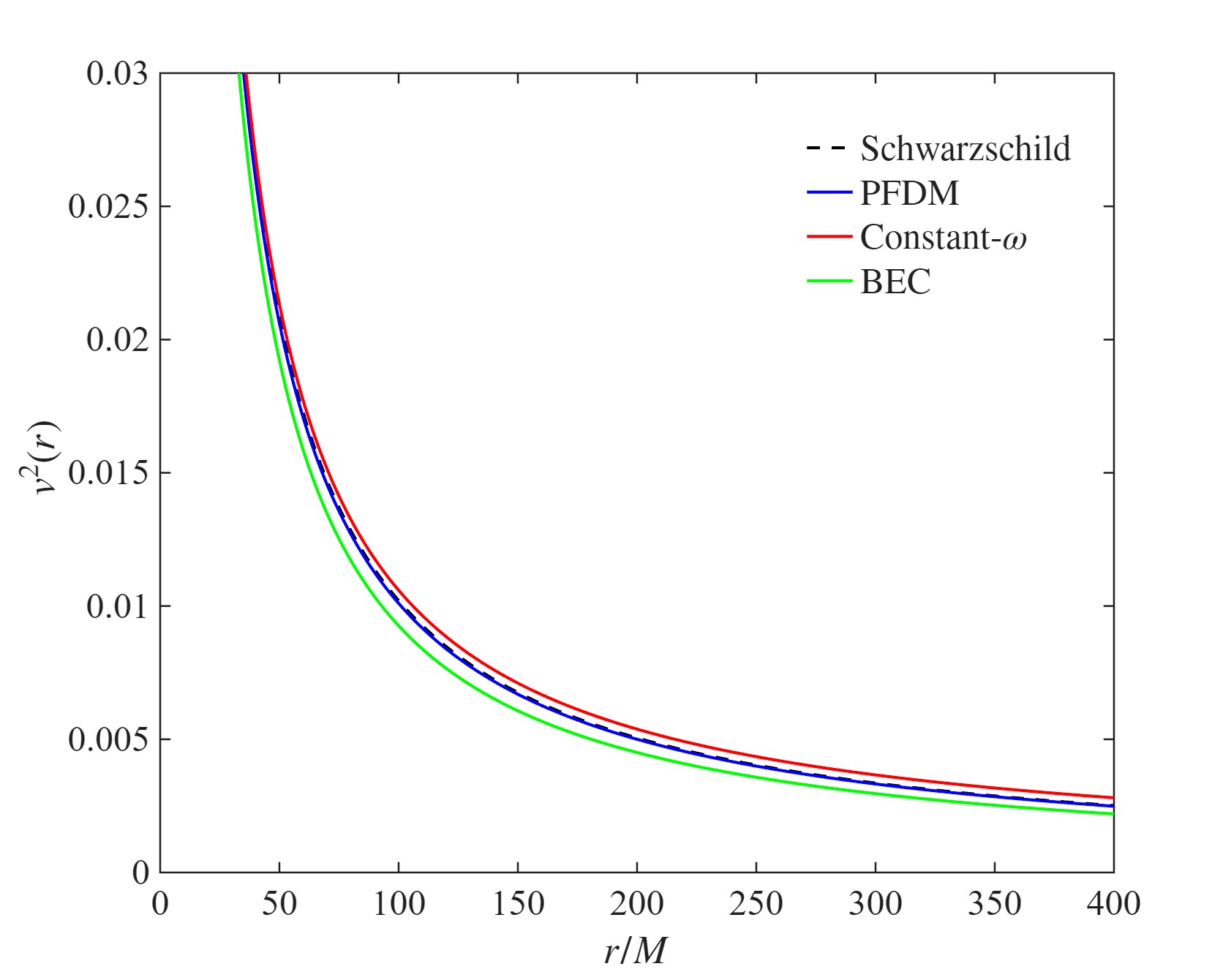}
    \caption{Radial dependency of the circular orbital velocity $v^2(r)$ across several dark matter models. }
\end{figure}
The circular orbital velocity features shown in Fig. 7 illustrate the influence of different dark matter hypotheses on the particle dynamics in a gravitational field. In both cases the velocity decreases with increasing radial distance which is consistent with the expected decrease of gravity at larger distances in relativistic spacetimes.The Schwarzschild solution is the reference case with the well known inverse radial dependence of the velocity.\\
The presence of dark matter produces a large deviations from this baseline. Among the models examined, the largest orbital velocities are found in the constant-omega scenario, which is attributed to a higher effective gravitational pull because of the pressure-dependent equation of state. The PFDM model also improves the velocity compared to the Schwarzschild case, but not as much. On the other hand, the BEC model provides relatively lower velocities, so that the change of the spacetime geometry is less noticeable. The differences are more evident in small radii where the gravitational field is stronger, while the curves tend to be closer to each other in larger distances.
The convergence means that the effect of dark matter decreases as you move away from the central object. The general trend suggests that the properties of dark matter, and in particular its equation of state, can have an effect on the orbital dynamics of test particles that is measurable.\\
From a physical standpoint, alterations in orbital velocity may affect observable characteristics, including accretion disk dynamics and emission processes in proximity to compact objects. Consequently, examining these velocity patterns offers a valuable foundation for investigating the characteristics of dark matter under intense gravitational settings \cite{nampalliwar2021modeling, rayimbaev2021dynamics}.\\
Figure 8 illustrates the physical feasibility of the PFDM spacetime by examining the energy conditions' behavior. The radial null energy condition is consistently positive over spacetime, signifying that the effective distribution of dark matter adheres to the fundamental positivity criterion for null observers. The dominant energy condition is satisfied for all examined values of $\epsilon$, indicating that the energy density prevails over the tangential pressure and that the matter content exhibits a physically acceptable behavior.\\
Conversely, the strong energy condition turns negative in proximity to the compact region $(r/M \sim 2)$, especially for elevated values of $\epsilon$ \cite{rubakov2008infrared, nojiri2003modified}. This infraction indicates the onset of repulsive gravitational phenomena caused by the dark matter surroundings adjacent to the black hole. This SEC violation is typically linked to unconventional effective matter distributions and altered gravitational behavior in strong-field conditions. As the radial distance grows, all energy-condition curves progressively converge to zero, signifying that the PFDM corrections become insignificant at considerable distances from the compact object, and the spacetime asymptotically approaches the Schwarzschild limit.
\begin{figure}[H]
    \centering
    \includegraphics[width=1\textwidth]{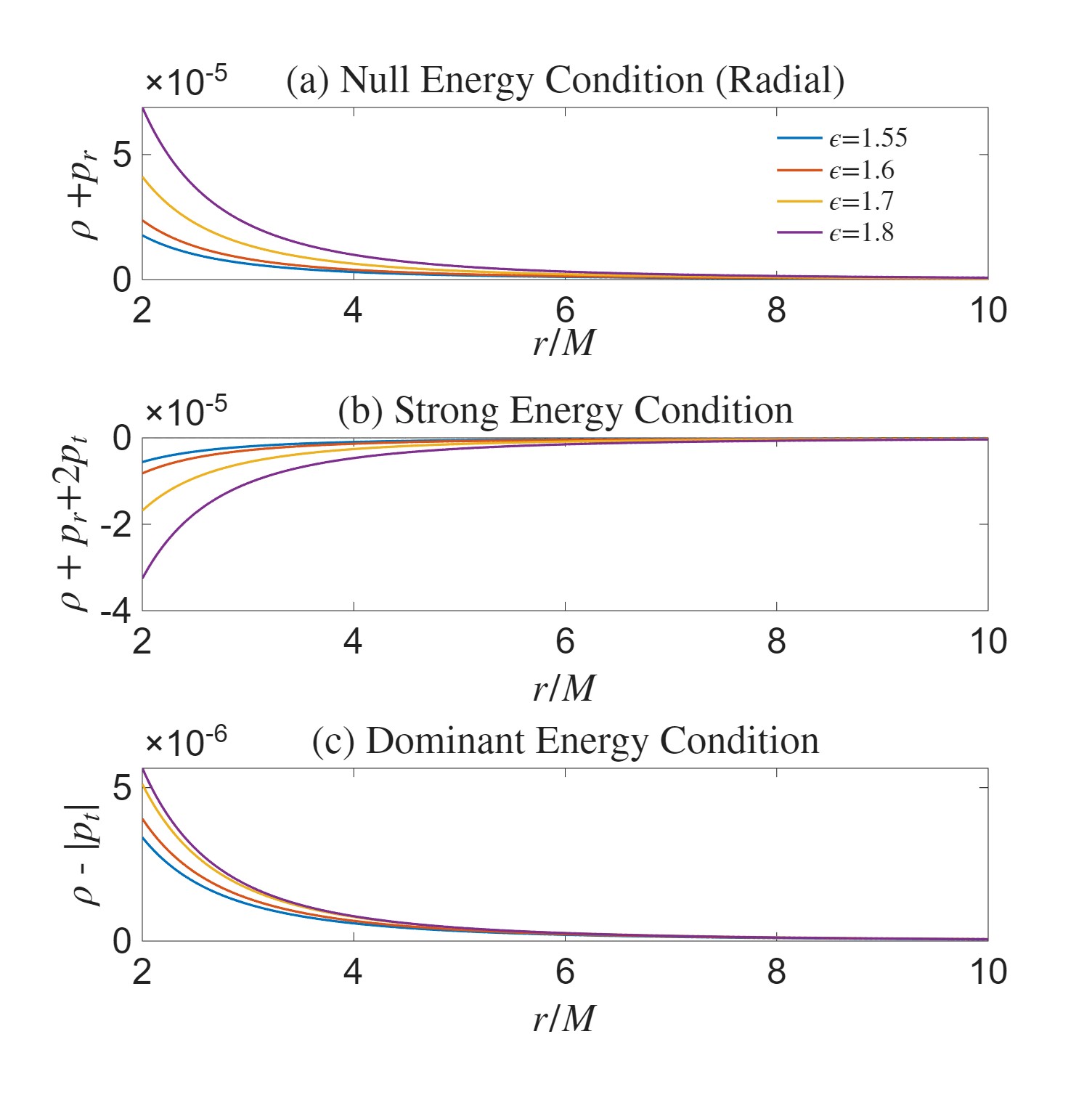}
    \caption{Energy conditions for the PFDM spacetime over many values of the parameter $\epsilon$. Panel (a) shows the radial null energy condition $\rho + p_r$, panel (b) represents the strong energy condition $\rho + p_r +2p_t$, and panel (c) depicts the dominant energy condition $\rho - |p_t|$ as function of the radial coordinate $r/M$.}
\end{figure}
Figure 9 depicts the dynamics of the energy conditions within the constant-$\omega$ dark matter environment encircling the black hole. The null energy condition is consistently positive across spacetime, signifying that the effective matter distribution adheres to the fundamental positivity criterion for physically plausible matter fields. Likewise, the dominant energy condition stays affirmative, indicating that the  matter-energy density above  pressure contribution and that causal energy transmission is maintained. \\
The strong energy criterion is fulfilled for the selected range of $\omega$, as evidenced by the positive values of $\rho + 3p$. This indicates that the gravitational field produced by the dark matter halo stays attracting and does not exhibit repulsive effects within the examined parameter range. Augmenting the EoS parameter $\omega$ amplifies the intensity of all energy-condition functions in proximity to the compact object, indicating that pressure effects intensify with bigger $\omega$. At extensive radial distances, all curves diminish toward zero, indicating that the dark matter contribution asymptotically declines and the spacetime progressively approaches Schwarzschild behavior far from the black hole.

\begin{figure}[H]
    \centering
    \includegraphics[width=1\textwidth]{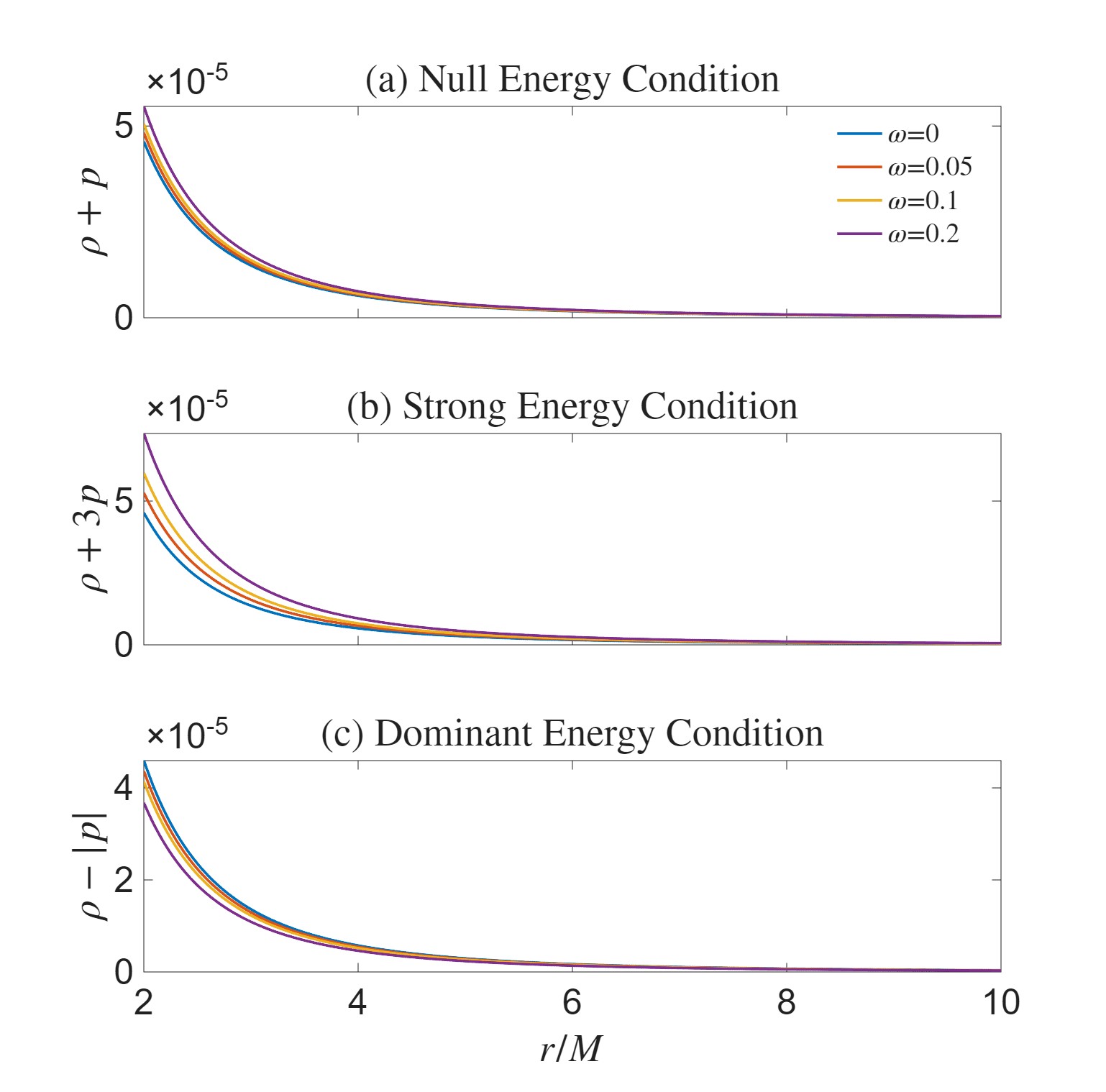}
    \caption{Energy conditions for constant-$\omega$ dark matter spacetimes over various values of the EoS parameter $\omega$. Panel (a) shows the null energy condition $\rho + p$, panel (b) represents the strong energy condition $\rho + 3p$, and panel (c) depicts the dominant energy condition $\rho - |p|$ as function of the radial coordinate $r/M$.}
\end{figure}

\begin{figure}[H]
    \centering
    \includegraphics[width=1\textwidth]{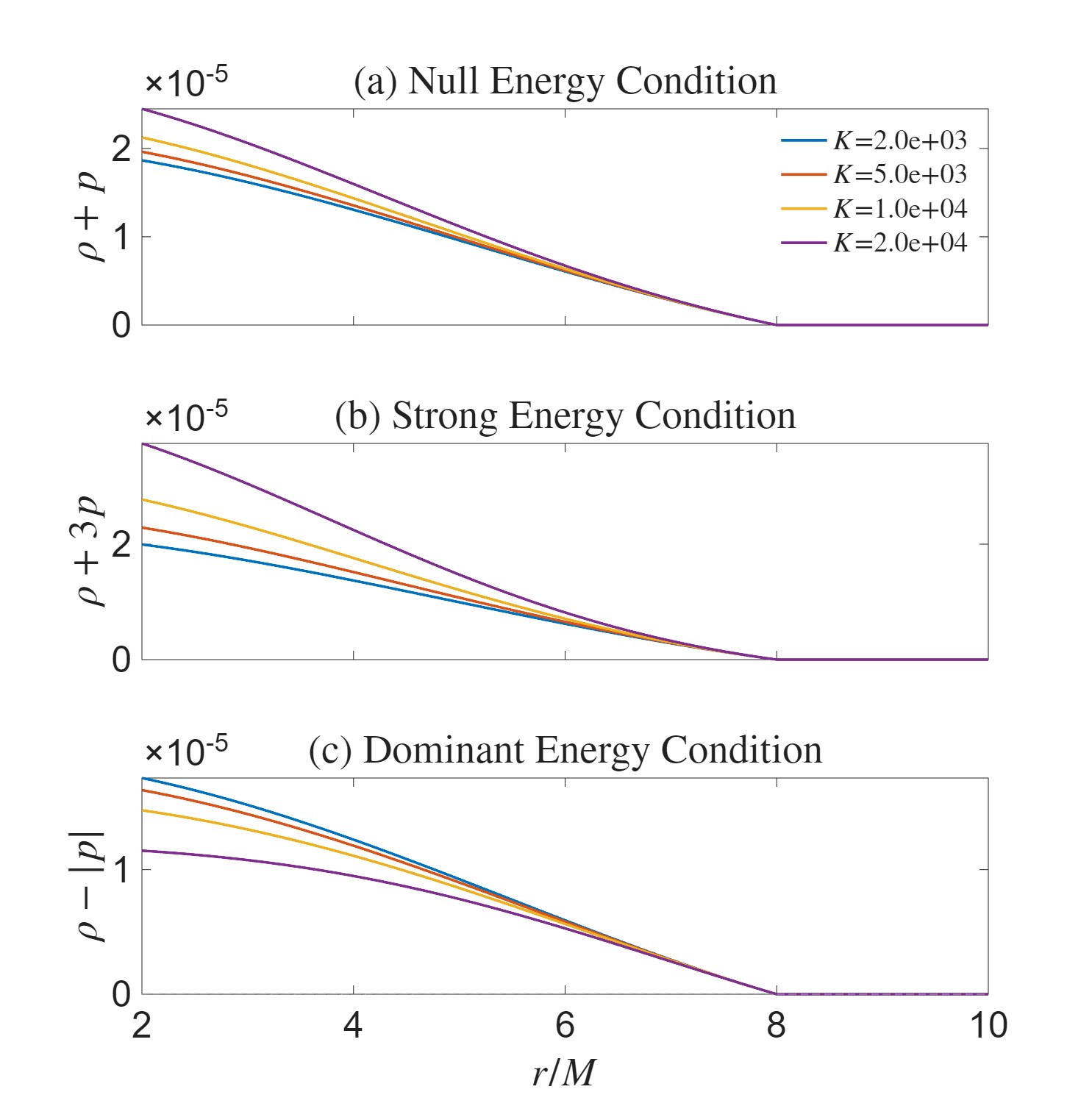}
    \caption{Energy conditions for the Bose Einstein Condensate (BEC) dark matter halo surrounding a black hole for different values of the interaction parameter K.}
\end{figure}
Figure 10 illustrates that the BEC dark matter distribution complies with the principal energy close to the black hole, signifying that the condensate functions as a physically permissible matter source. The affirmative behavior of the null energy condition verifies that the physical energy density stays non-negative when integrated with the pressure component. Similarly, the dominant energy condition remains positive across the spacetime, indicating that the energy density prevails over the effective pressure and guarantees causal transmission of matter-energy flow.\\
The strong energy condition remains positive for all examined values of the interaction strength $K$, indicating that gravitational field produced by the condensate maintains an attractive character. As the parameter $K$ increases, the pressure resulting from bosonic self-interactions intensifies, resulting in greater magnitudes of the energy-condition functions around the compact object. This indicates the growing impact of quantum pressure effects within the condensate halo.\\
All curves are decreasing continuously with the increase of the radial distance, asymptotically approaching zero, which indicates that the effect of BEC halo decreases far from the black hole, while the spacetime smoothly tends to the Schwarzschild limit at large distances. The results confirm the physical consistency and stability of the BEC dark matter model in describing black hole environments in modified dark matter scenarios.

\section*{Conclusions}
We studied the impact of different dark matter environments on the geometric and dynamical properties of black holes by considering three theoretical models: Perfect Fluid Dark Matter (PFDM), Constant-$\omega$ dark matter and Bose Einstein Condensate (BEC) dark matter. The investigation was carried out by studying the horizon structure, the raduis of the photon sphere, the innermost stable cicular orbit(ISCO), the circular orbital motion and the energy conditions. \\
Our results show that the dark matter influences the spacetime geometry around the black hole and causes observable variation from standard Schwarzschild case. The dark matter parameter $\epsilon$ in the PFDM model causes substantial changes in the photon sphere and ISCO radius, which indicates that the distribution of matter around the black hole plays a remarkable role in the dynamics of test particles and the structure of accretion disks. The increase of $\epsilon$ is generally pushing the stable orbit outward, showing the enhanced gravitational effect of halo.\\
In the Constant-$\omega$ dark matter scenario the EoS parameter $\omega$ is crucial to determine the orbital structure and horizon dynamics. By modifying the value of $\omega$, we can alter the effective gravitational attraction around the compact object, leading to profound changes in the ISCO radius and the orbital velocity profiles. The study shows a significant effect of pressure forces of the dark matter fluid on the stability of a circular motion in the vicinity of the black hole.\\
In the BEC dark matter model the bosonic self interactions leads to quantum pressure which produces smooth and stable modifications of the spacetime geometry. The interaction parameter $K$ not only affects the orbital dynamics but also the energy distribution around the black hole. The analysis of the energy condition indicates that the BEC halo is a physically acceptable source of matter within the range of the parameters considered.\\
The physical consistency of the three dark matter concepts is also demonstrated by the study of the energy conditions. For suitable choices of parameters, the null, strong and dominating energy conditions are satisfied implying that the effective matter distributions remain stable and gravitationally attractive near the black hole. Moreover, all dark matter effects tend to vanish at large radial distances, as the spacetime approaches the Schwarzschild limit asymptotically.\\
This study's results highlight the dark matter's significant influence on black hole observables and orbital characteristics. Such discrepancies could be useful theoretical indicators for future astrophysical findings on accretion disks, black hole shadows, gravitational lensing and galactic rotation curves. This analysis offers a useful starting point for understanding black holes in realistic dark matter settings, and may inform future work connecting compact objects to models of cosmic dark matter.


\bibliography{refs}
\end{document}